# Evaluation of Spin Waves and Ferromagnetic Resonance Contribution to the Spin Pumping in Ta/CoFeB Structure

Mahdi Jamali[a], Angeline K. Smith[a], Hongshi Li[b] and Jian-Ping Wang [a]

[a]Department of Electrical and Computer Engineering, University of Minnesota, 4-174 200 Union Street SE, Minneapolis, MN 55455, USA

[b]Department of Chemical Engineering and Materials Science, University of Minnesota, Minneapolis, MN 55455

The spin waves and ferromagnetic resonance (FMR) contribution to the spin pumping signal is studied in the Ta/CoFeB interface under different excitation bias fields. Ferromagnetic resonance is excited utilizing a coplanar waveguide and a microwave generator. Using a narrow waveguide of about 3 μm, magnetostatic surface spin waves with large wavevector ($k$) of about 0.81 μm$^{-1}$ are excited. A large $k$ value results in dissociation of spin waves and FMR frequencies according to the surface spin wave dispersion relation. Spin waves and FMR contribution to the spin pumping are calculated based on the area under the Lorentzian curve fitting over experimental results. It is found that the FMR over spin waves contribution is about 1 at large bias fields in Ta/CoFeB structure. Based on our spin pumping results, we propose a method to characterize the spin wave decay constant which is found to be about 5.5±1.27 μm in the Ta/CoFeB structure at a bias field of 600 Oe.

Electronic address: jpwang@umn.edu

Spin pumping as a mechanism of spin current generation due to dynamical magnetization states has absorbed vast attention among researchers recently [1–4]. Pumping of spin current has been successfully demonstrated from a magnetic layer into metallic channels [2,5–7], semiconductors [8,9], insulators [10] and more recently into topological insulators [11,12]. In the spin pumping experiment, the magnetization dynamics is usually excited utilizing a microwave source [1,5,13]. The injected spin current into the nonmagnetic channel is detected by means of the inverse spin Hall effect of the nonmagnetic channel [14,15]. Typically, ferromagnetic resonance (FMR) is intended to be the main part of magnetization dynamics in the spin pumping experiment; however, due to the geometrical effect of the coplanar waveguide [1] or the magnetic layer itself [10,16,17], spin waves are often excited as well. Geometrical effect of the waveguides refers to the finite size of the waveguide compared to the magnetic layer that results to a non-uniform excitation. Geometrical effect of the magnetic layer refers to the spin waves confinement that results in the spin waves quantization and formation of spin waves standing waves. Estimation of spin waves and FMR contribution to the spin pumping signal is crucial knowing that the spin Hall angle of the nonmagnetic material can be extracted from the amplitude of the spin pumping signal [5,18,19]. In the spin pumping experiments with ferromagnetic metallic layers such as CoFeB[5] or NiFe[4], spin waves and FMR frequencies are close. Moreover, due to broadening of the FMR linewidth which is enhanced by the nonmagnetic heavy metal like Ta or Pt, only a single resonant frequency can be observed in the spin pumping experiment[5]. Most of the previous works on the investigation of spin wave contribution to the spin pumping are based on using wide waveguides that result in mixing of spin waves and FMR signals[20] and make distinction of the them very complex. Multiple eigenfrequencies in spin pumping experiments are mostly reported in the experiments based on the YIG magnetic oxide[21,22].

In this letter, we have estimated the relative intensity of spin wave and FMR contributions to the output spin pumping signal in Ta/CoFeB bilayer metallic systems.

The output spin pumping signal is the electromotive force generated by the inverse spin Hall effect of the Ta channel acting on the spin current generated from the magnetization dynamics of the CoFeB layer. By employing a narrow waveguide, surface spin waves with a wavevector of about 0.81 μm$^{-1}$ are excited in the CoFeB layer resulting in a large difference between the FMR and spin waves frequencies. In addition, based on our results, the authors have proposed a method to extract the spin waves characteristic decay length in ferromagnetic metallic system which is found to be about 5.5±1.27 μm in the Ta/CoFeB structure at a bias field of 600 Oe. In this study, FMR refers to the nonpropagating magnetization dynamical mode. Due to the large linewidth of spin pumping signal, it also includes the quasi-static spin waves mode with very small wavevector. The propagating modes of magnetization dynamics are spin waves that have a sizable wavevector. The spin waves have different resonant frequencies compared to the FMR mode. The spin waves modes present in our experiments are mostly magnetostatic surface spin waves modes where the bias magnetic field and the spin waves wavevector are in the plane of the magnetic film and normal to each other[5,23,24].

Exchange-coupled spin waves with a large wavevector (short wavelength) in metal/magnetic insulator structure have been studied using parametric excitation or narrow waveguides [25–27]. Ferromagnetic metallic layers utilized in the spin pumping experiments are usually very thin (~ 10 nm) and their damping constant is much larger than magnetic insulators. Since parametric spin pumping is not easily achievable in a metallic ferromagnet, to obtain a sizable difference between the frequency of FMR and spin waves, a narrow microwave waveguide must be utilized.

Fig. 1(a) shows a schematic of the device structure and measurement setup. Initially, Ta (5 nm) is sputter deposited utilizing a 6-target Shamrock sputtering system with a built-in argon ion miller on a thermally oxidized Si substrate with a $SiO_2$ thickness of about 300 nm. The Ta film is patterned using photolithography into rectangular shapes with a size of 200 μm× 50 μm using negative resist and subsequent argon ion milling. Next, by lift-off process, $Co_{20}Fe_{60}B_{20}$ (10 nm) with a size of 30 μm×50 μm is placed on top of the Ta channel. The surface of Ta layer is slightly etched (~0.4 nm) before CoFeB deposition to provide a fresh interface between the Ta and CoFeB. The samples are field-annealed in a vacuum system with a base pressure of less than $1\times10^{-6}$ Torr in the presence of a magnetic field of 0.4 T and temperature of 300°C for 2hrs. Magnetization dynamics is excited using an asymmetric coplanar waveguide in the GS form and a sinusoidal microwave source. The waveguide is isolated from the magnetic layer and Ta channel by $SiO_2$ (50 nm) that is deposited by electron beam evaporation. An optical micrograph of the fabricated device is given in Fig. 1(b). The waveguide has signal and ground lines with widths of 3 and 9 μm, respectively, and the spacing between them is 3 μm.

Upon excitation of the magnetization dynamics by the rf-field generated from the coplanar waveguide, spin current ($J_s$) is pumped into the Ta layer (in z-direction). Both FMR and spin waves are excited by the waveguide and contribute to the pumping of the spin current. Magnetic field is applied along the x-direction during the measurements. Strong spin-orbit interaction of the Ta layer translates $J_s$ into a charge current $J_c$ due to the inverse spin Hall effect (ISHE). The electric field induced by the ISHE could be written as[22]:

$$E_{ISHE} \propto J_s \times \sigma \qquad (1)$$

where $J_s$ is the spin current injected from CoFeB in Ta and σ is the spin polarization vector of the spin current defined by the bias magnetic field. Magnetization dynamics in the CoFeB generates nonequalibiruim polarized electron in the adjacent nonmagnetic layer[28]. The pumped spin current results in additional damping of the magnetic layer itself. An electromotive force is generated across the Ta channel in the *y*-direction that can be detected by a nano-voltmeter. The spin-orbit interaction is responsible for the inverse spin-Hall effect (ISHE) and is a process that converts a spin current into an electric voltage. The strong spin-orbit interaction in heavy metals like Pt and Ta[23] allows observation of the ISHE at room temperature.

The frequency spectra of the output dc-voltage at ±130 Oe is presented in Fig. 2(a). As seen, the output voltage polarity is altered by changing the magnetic field polarity which is consistent with the spin pumping experiments reported by other groups[4,5]. In most of the previous spin pumping works based on metallic ferromagnets like NiFe or CoFeB, only a single resonant peak is observed [5,6,29]; however, in this experiment, three frequencies are present in the output voltage spectra. The main frequency occurs at 6.2 GHz which is associated with FMR excitation while the higher frequencies of 6.7 and 8.3 GHz are correlated with spin waves excitation in the magnetic layer. Due to the narrow width of the waveguide, spin waves with large wavevectors can be excited. Since the magnetization of CoFeB is in-plane and the magnetic field is applied along *x*-direction while spin waves propagation is along the *y*-direction, magnetostatic surface spin waves (MSSW) are excited in the CoFeB[23,24]. It is well known that MSSW shows nonreciprocal behavior for opposite field polarities[23,30]. The non-reciprocity of MSSW is indeed observed in our experiments for the spin pumping signal at positive and negative fields due to asymmetric coplanar waveguides. The difference between the amplitude of the spin pumping signal at positive and negative fields is less than 10% and we have safely neglected it in our calculation. The frequency spectra of spin pumping is shown in Fig. 2(b)-(c) for ±260 and ±390 Oe, respectively. The spin pumping resonant frequency corresponding to FMR is shifted to 8.3 and 10 GHz for the bias magnetic fields of 260 and 390 Oe. The second resonant frequency is shifted to 8.7 GHz at 260 Oe and it is merged with the FMR peak at 390 Oe. This resonant peak could present the nonhomogeneous magnetization excitation that disappears at large bias fields due to complete magnetization saturation along the field direction. The third peak that corresponds to MSSW is changed to 10.1 and 11.5 GHz at the fields of 260 and 390 Oe, respectively. There are also contributions from the anisotropic magnetoresistance (AMR) and/or the anomalous Hall effect (AHE) of the magnetic layer (CoFeB) in the output voltage. Both AMR and AHE have the form of asymmetric Lorentzian functions and can be isolated from the spin pumping signal[19,31,32].

The effect of the excitation amplitude on the spin pumping frequency spectra is shown in Fig. 2(d) for the bias magnetic field of -200 Oe. By increasing the excitation amplitude from 0 dBm to 7 dBm, the amplitude of the spin pumping increases accordingly. The amplitude of FMR and MSSW peaks at 0 dBm (1 mW) excitation are 1.5 and 1.7 µV while for the excitation power of 7 dBm (5 mW), they are changed to 7.8 and 9.4 µV, respectively. Moreover, the resonant peak positions from spin pumping for the first two peaks at 5.9 and 6.6 GHz are the same upon increasing the power from 0 to 7 dBm showing negligible nonlinear effect due to the input excitation power. Only the third peak at 8.6 GHz shows slight red-shift down to 8.3 GHz by the increasing of the input power that could be associated with the nonlinear behavior of spin waves at large input power. This is expected since narrow coplanar waveguides can generate large rf-fields at high input power.

Fig. 2(e) is a schematic image of the device showing the profile of the magnetization excitation in our structure. FMR is mostly excited in the CoFeB layer located under the waveguide. Surface spin waves are also excited at the same time in the CoFeB which propagate toward left and right with a wavevector of $K_{sw}$. Due to decay of the spin waves along the propagation direction in the magnetic layer, the injected spin current by the spin waves into the Ta layer is also non-uniform and decay accordingly. In Fig. 3(a), the resonant frequency of the first peak (that is merged with the second peak at high bias field) and the third peak (that is the second peak at high field) in the spin pumping spectrum corresponding to FMR and MSSW at different bias magnetic fields are demonstrated. Both FMR and MSSW peaks show behavior that is consistent with their dispersion relation. FMR dispersion follows the Kittel formula: $f = \frac{\gamma}{2\pi}\sqrt{(H+H_a)(H+H_a+\mu_0 M_{eff})}$ where $\gamma$ is the gyromagnetic ratio, $M_{eff}$ is the effective saturation magnetization of thin film, and $H_a$ accounts for shape/crystalline anisotropy. Upon curve fitting of the Kittel formula over the FMR data, the corresponding value for $\gamma$ and $M_{eff}$ are found to be $2.9\times10^5$ m.A$^{-1}$.s$^{-1}$ and $1.3\times10^6$ A/m, respectively. Utilizing the relation $\gamma = \frac{g\mu_B}{\hbar}$, a Landé g-Factor ($g$) of about 2.6 is obtained for the CoFeB thin film. This value is slightly larger than what is reported by another group[33] for the perpendicular CoFeB thin film. One possible reason could be because the spin pumping effect in the in-plane film enhances the effective damping and increases the effective spin-orbit coupling of the CoFeB layer at the interface with the Ta layer. Magnetostatic surface spin waves also known as Damon-Eshbach spin waves[24] are defined by the dispersion relation: $f = \frac{\gamma}{2\pi}\sqrt{(H+H_a)(H+H_a+\mu_0 M_{eff})+(\frac{\mu_0 M_s}{2})^2(1-e^{-2k_{sw}d})}$ where $d$ is the magnetic layer thickness (10 nm). From the curve fitting of MSSW dispersion relation over the experimental data, the spin wave wavevector is calculated to be about 0.81 µm$^{-1}$ corresponding to a wavelength of 7.8 µm.

We has also performed a micromagnetic simulation to understand the origin of the peaks that are present in the spin pumping spectra. One dimensional micromagnetic simulation is performed with a cell size of 25 nm×200 μm×10 nm using OOMMF package[23,34,35]. Magnetization dynamics are excited by a Gaussian field pulse of 50 ps for different bias fields. Fig. 3(b) shows the spin wave wavevector spectrum extracted from the simulation after 2 ns from the pulse field excitation for the magnetic field of 200 Oe. The wavevector is calculated using the fast Fourier transform of spatial distribution of magnetization dynamics. As seen in Fig. 3(b), the main wavevector of the surface spin waves happens at 0.98 μm$^{-1}$ corresponding to the wavelength of about 6.5 μm which is close to the experimental value of 7.8 μm. In addition, the second peak witnessed in the experimental results is not observed in the simulation confirming that it is due to nonhomogeneous magnetization excitation.

The relative intensity of spin waves and FMR contribution to the spin pumping is calculated based on the area of the Lorentzian curve corresponding to FMR and spin waves. The curve fitting of the Lorenzian curve over the experimental results are presented for the bias fields of -600 and 80 Oe in Fig. 4(a) and Fig. 4(b), respectively. Having the Lorentzian curve of $V_{FMR/SW} = \dfrac{A_{FMR/SW}}{(f-f_0)^2 + \eta^2}$, the intensity of FMR and spin wave contribution to the spin pumping are given by $I_{FMR/SW} = \sqrt{\dfrac{\int_{-\infty}^{+\infty} V_{FMR/SW}^2(f)df}{\Delta f_{FMR/SW}}}$ . In this formula $\Delta f_{FMR/SW}$ is the FMR/SW frequency linewidth. At a bias field of -600 Oe, the ratio of the FMR to spin wave contribution to the spin pumping signal is about 1.0. At low bias fields of 80 Oe, this ratio drops to 0.8 Therefore, in ferromagnetic metallic layers with narrow waveguides, the spin wave contribution to the spin pumping signal is equally important compared to the FMR and it must be considered. This is very significant especially when the spin Hall angle is estimated from the spin pumping signal.

The ratio of spin waves to FMR contribution could be utilized to estimate the spin wave decay length in the magnetic layer once a heavy metal is in contact with the magnetic metallic layer. This is especially useful knowing that the spin wave decay constant is much shorter in the presence of the heavy metal and direct characterization of spin waves is difficult. The injected spin current into the nonmagnetic layer is proportional to $\sin^2(\theta)$[5,6] where $\theta$ is the magnetization precession cone angle. Assuming that the FMR precession cone angle is $\theta_0$ which happens only in the area under the waveguide, the spin waves propagate toward left and right with an exponential decay constant of $\Lambda$. The FMR precession angle can be derived from the below formula knowing the input rf-field ($h_{rf}$).

$$\theta = \frac{h_{rf}}{\Delta H \sqrt{1 + \frac{(H - H_r)(H + H_r + 4\pi M_s)}{\Delta H 4\pi M_s}}} \quad (1)$$

Here $H_r$ is the resonant field and $\Delta H$ is the FMR linewidth. Thus, the maximum precession cone angle equals $\theta = \frac{h_{rf}}{\Delta H}$. Knowing that $\Delta H = \Delta f \frac{dH}{df}$, $\Delta H$ can be obtained from the spin pumping frequency spectra utilizing the relation $\Delta H = \frac{\Delta f}{\gamma \sqrt{1 + (\frac{\gamma \mu_0 M_{eff}}{4\pi f})^2}}$ [5,36].

Assuming that the spin wave precession angle is the same as FMR at the boundary of the waveguide ($\pm L/2$), the spin wave precession angle could be obtained from $\theta = \theta_0 e^{-\frac{x - L/2}{\Lambda}}$ for $x \geq \frac{L}{2}$ and $\theta = \theta_0 e^{-\frac{x + L/2}{\Lambda}}$ for $x \leq \frac{-L}{2}$ as shown in Fig. 5. This estimation is quite accurate once the spin waves and FMR frequencies are close. The precession cone angle can be extracted utilizing Eq. (1). At the bias field of 600 Oe, the spin waves and FMR resonant frequencies are 13.6 and 12.36

GHz, respectively. Moreover, the spin waves and FMR frequency linewidths are 0.75 and 0.6 GHz, respectively. The precessional cone angle is proportional to the inverse of the frequency linewidth. Therefore, there is about 0.23% difference in the cone angle of spin waves and FMR at the bias field of 600 Oe in our experiment.

The ratio of FMR to spin wave contribution could be estimated by this formula:

$$\frac{I_{FMR}}{I_{SW}} = \frac{\int_{-L/2}^{L/2} \sin^2(\theta_0) dx}{2\int_{L/2}^{\infty} \sin^2(\theta) dx} \qquad (2)$$

This ratio can be solved numerically as a function of $\Lambda$. At the bias field of 600 Oe, the spin wave decay constant is found to be about 5.5±1.27 μm. The curve fitting of Lorentzian curve over the experiment results has less than 5% error. According to our previous work, the Gilbert damping constant increases by a factor of about 2.5 in Ta/CoFeB bilayer structure compared to the CoFeB layer[5]. Assuming the spin waves decay constant is proportional to the Gilbert damping, the spin waves decay length in a CoFeB thin film is estimated to be about 13.75 μm at large bias fields.

In summary, the spin wave contribution to the spin pumping signal in Ta/CoFeB bilayer is studied experimentally. Using a narrow waveguide of 3 μm width, magnetostatic surface spin waves with a wavevector of about 0.81 μm$^{-1}$ are excited that results in large dissociation of spin waves and FMR resonant frequencies. Based on the ratio of spin waves to FMR contribution to the spin pumping signal, a method is proposed to estimate the spin wave decay constant in the bilayer heavy metal/magnet structure. Our experimental results and proposal pave the way in better understanding of the spin wave contribution to the spin pumping signal and it could be utilized to characterize spin waves in metallic systems by means of spin pumping. Additionally, this shows a significant contribution of spin waves to the spin pumping signal for narrow co-planar waveguides.

This is critical for understanding and correct interpretation of results when using spin pumping experiments for determination of material parameters such as spin Hall angles.

Acknowledgement

This work was partially supported by the C-SPIN center, one of six STARnet program research centers, and National Science Foundation Nanoelectronics Beyond 2020 (Grant No.NSF NEB **1124831**). Parts of this work were carried out in the Minnesota Nano Center which receives partial support from NSF through the NNIN program.

Figure Captions

FIG. 1.  (a) A schematic of spin pumping characterization device in Ta/CoFeB bilayer structure. Magnetization dynamics is excited using an asymmetric coplanar waveguide and the output electromotive force is characterized by a nanovoltmeter. (b) An optical micrograph of the actual fabricated device where individual layer is labeled.

FIG. 2.  The spin pumping frequency spectra at the bias field of (a) ±130 Oe, (b) ±260 Oe, and (c) ±390 Oe. (d) The spin pumping spectra characterized at a bias field of -200 Oe and for the excitation power of 0, 2, 4, and 7 dBm. (e) A schematic showing how spin waves and FMR excite and contribute to the pumping of spin current. The spin current indicating with down arrows are injecting into the Ta channel by both FMR and spin waves.

FIG. 3.  (a) The FMR and magnetostatic surface spin waves resonant frequencies obtained at different bias fields from the spin pumping experiment. (b) The spin waves wavevector

for a bias field of 200 Oe and an excitation field pulse of 50 ps obtained from micromagnetic simulation.

FIG. 4. (a) The curve fitting of Lorentzian function over the experimental results for bias fields of (a) -600 Oe and (b) -80 Oe.

FIG. 5. A schematic showing the magnetization precession cone angle distribution under and away from the coplanar waveguide.

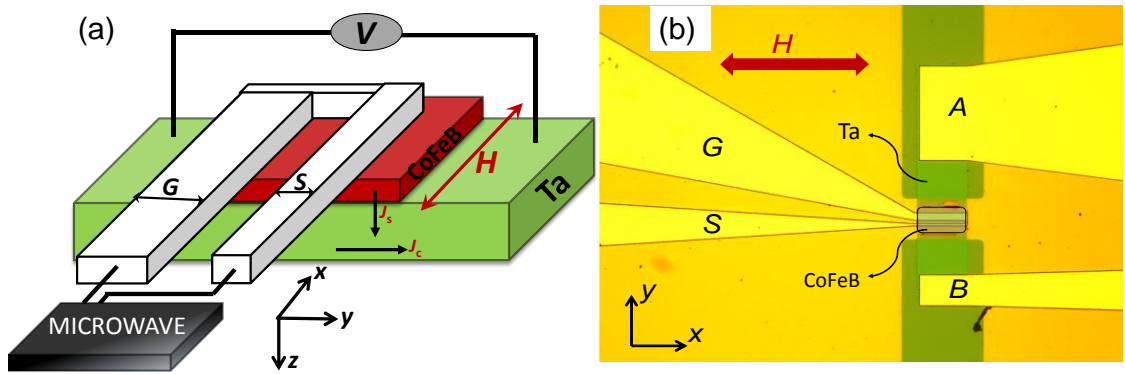

Figure 1

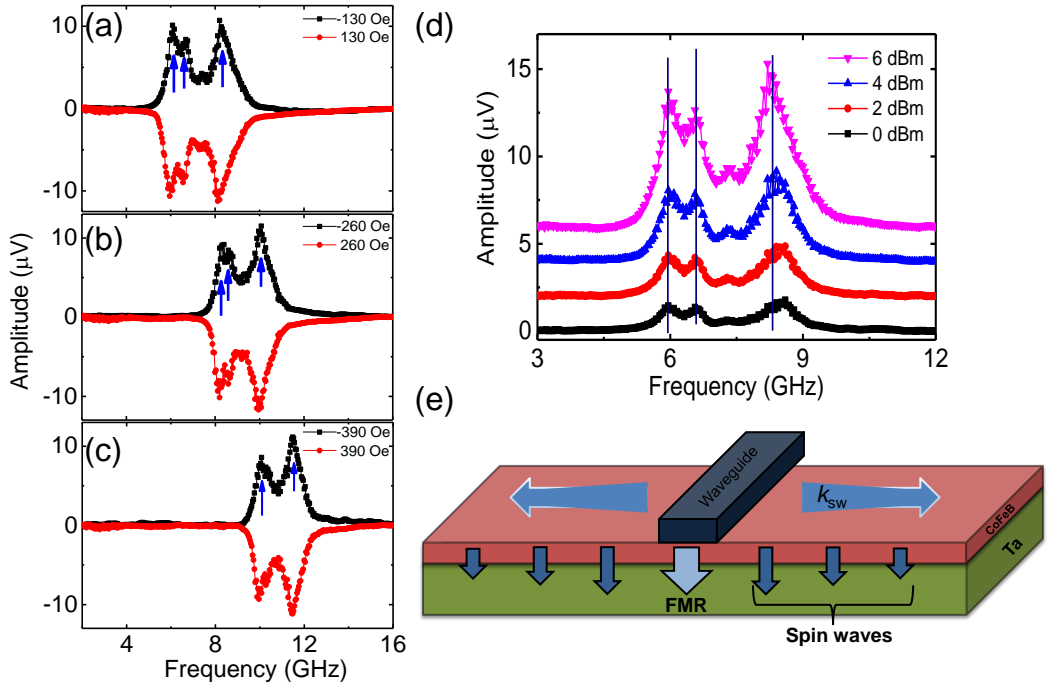

Figure 2

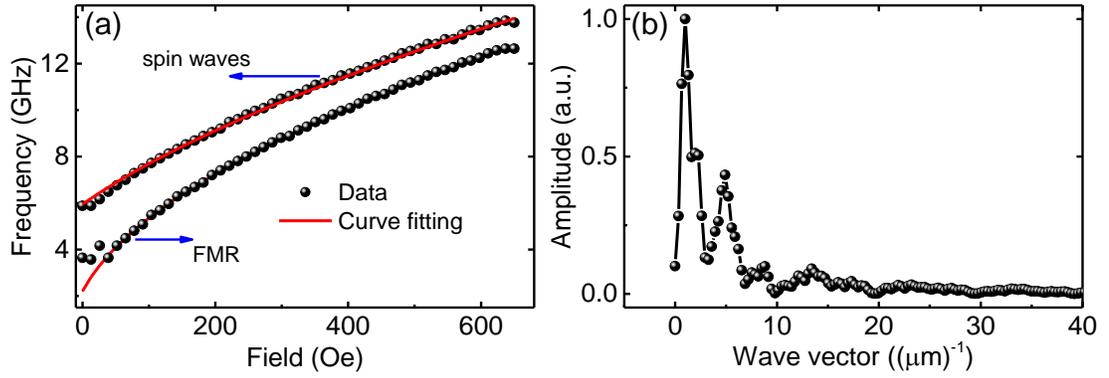

Figure 3

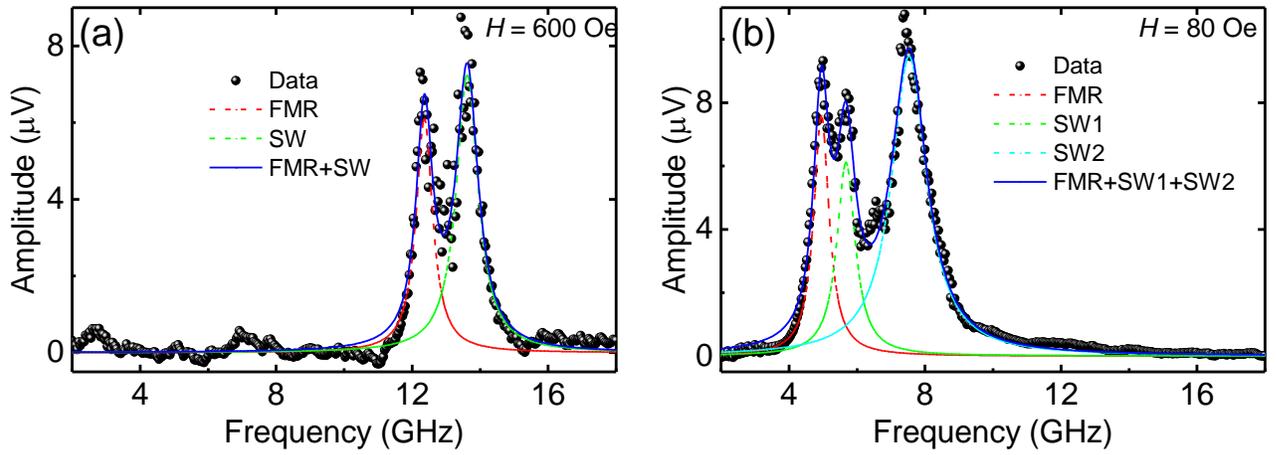

Figure 4

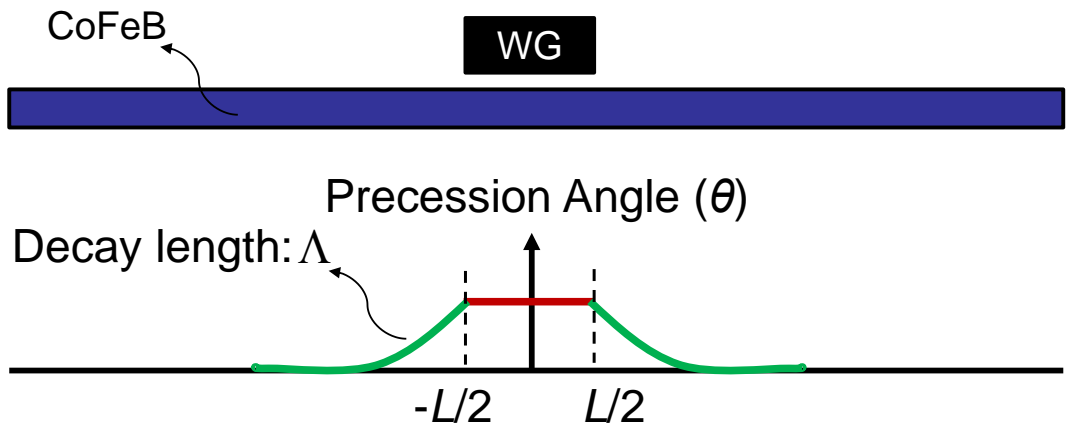

Figure 5